\documentclass[proceedings]{JHEP3}
\usepackage{eurosym}
\usepackage{amssymb}
\usepackage{amsfonts}
\usepackage{braket}
\usepackage{amsmath,epsfig}
\usepackage{graphicx}
\usepackage{multirow}
\usepackage{float}
\usepackage{xcolor}
\newbox\mybox

\newcommand\fverb{\setbox\mybox=\hbox\bgroup\verb}
\newcommand\fverbdo{\egroup\medskip\noindent\fbox{\unhbox\mybox}\ }
\newcommand\fverbit{\egroup\item[\fbox{\unhbox\mybox}]}
\conference{Complex topological soliton with real energy in particle physics}
\abstract{We summarise the procedure used to find the classical masses of Higgs particle, massive gauge boson and t'Hooft-Polyakov monopole in non-Hermitian gauge field theory. Their physical regions are explored, and the mechanism of the real value of the monopole solution is analysed in different physical regions.}

\title{Complex topological soliton with real energy in particle physics}
\author{Takanobu Taira \\
Department of Mathematics, City, University of London,\\
Northampton Square, London EC1V 0HB, UK \\
E-mail: a.fring@city.ac.uk, takanobu.taira@city.ac.uk}

\input{tcilatex}

\begin{document}
\section{Introduction}\label{Section: Ch3 Sec0}
    Quantum field theory is a key tool to analyse particle physics. The most modern physical description of the fundamental particle interaction is described by the model called the Standard Model. However, the model possesses several problems such as incompatibility with the general relativity, hierarchy problem, etc. Therefore, it is an active area to extend the standard model. 
     
    Recently a growing number of research papers started exploring the non-Hermitian extension of the Standard Model \cite{AlexandreNoether,AlexandreGoldtone,MannheimGoldstone,Millington2019a,Alexandre2019,AlexandreHiggs,Alexandre_Ref_on_ill_defined_Dyson_map,FringGoldstone,fring2020massive,fring2019pseudo,fring2020t,AlexandreSusy,chernodub2021spontaneous,chernodub2021ir}. We have contributed to this development by analysing the Goldstone theorem \cite{fring2019pseudo,FringGoldstone}, The Higgs mechanism \cite{fring2020massive} and t'Hooft-Polyakov monopoles \cite{fring2020t}. The classical masses of Higgs particles, massive gauge boson and monopoles were analysed. However, a detailed analysis of their intersecting physical regions and the mechanism of the real value of the energy of monopole was not explored. The main aim of this contribution is to fill this gap.
    
    There are two separate mechanisms that guarantee the real value of the particle masses in question. First, the masses of Higgs particles are given by a non-Hermitian mass matrix $M$. Assume that the matrix possess anti-linear symmetry, which we refer to as $\mathcal{
    PT}$ symmetry, that satisfies $[\mathcal{PT},M]=0$, $Mv=\lambda v$, $\mathcal{PT}v=e^{i\theta}v$, where $\{v,\lambda\}$ are eigenvectors and eigenvalues of the mass matrix. From this, it is trivial to show that the eigenvalues are real
    \begin{eqnarray}
    \mathcal{PT}M v_i &=& \mathcal{PT}\lambda_i  v_i = \lambda_i^* \mathcal{PT} v_i = \lambda_i^* e^{i\theta_i}v_i,\nonumber\\
    \mathcal{PT}M v_i &=& M\mathcal{PT} v_i = M e^{i\theta_i} v_i = \lambda_i e^{i\theta_i}v_i.\nonumber
    \end{eqnarray}
    It was shown in \cite{FringGoldstone} that this $\mathcal{PT}$ symmetry is related to the $\mathcal{CPT}$ symmetry of the field-theoretic action.

    On the other hand, the classical energy of the soliton solution is found by inserting the solution into the Hamiltonian $E=H[\phi] = \int d^3 x \mathcal{H}(\phi)$. Therefore, the techniques from $\mathcal{PT}$ symmetric quantum mechanics shown above can not be applied. 
    
    We will show below that the energy of the soliton solutions are real when the three conditions stated below holds. Therefore they are \textit{sufficient} conditions to guarantee the real value of particles in the model. However, we do not claim that these are \textit{necessary} conditions.
    
    Let $\{\phi_1, \phi_2\}$ be a set of distinct (or identical) solutions to the equations of motion $\delta \mathcal{L} / \delta \phi - \partial_\mu (\delta \mathcal{L}/ \delta \partial_\mu \phi)=0$, where $\mathcal{L}(\phi)$ is the field-theoretic Lagrangian density. The classical energies of the solution are given by inserting the solution into the Hamiltonian, $E_i=H[\phi_i]=\int d^3 x \mathcal{H}(\phi_i)$, for $i\in \{1,2\}$. The classical mass of the solution $\phi_1$ and $\phi_2$ are real if there exist some anti-linear symmetry $\mathcal{CPT}$ (note that is it not the standard $\mathcal{CPT}$ symmetry in quantum field theory) such that three conditions are satisfied: 
    \begin{enumerate}
        \item $\mathcal{CPT}: \mathcal{H}[\phi (x)]\rightarrow \mathcal{H}[\mathcal{CPT}\phi(x)]= \mathcal{H}^\dagger [\phi (-x)]$. \label{Reality condition 1}
        \item $\mathcal{CPT}: \phi_1 (x)\rightarrow \phi_2(-x)$.\label{Reality condition 2}
        \item $H[\phi_1]=H[\phi_2]$.\label{Reality condition 3}
    \end{enumerate}
    If two solutions are identical $\phi_1 = \phi_2$, then the above condition reduces to the reality condition of the soliton solution already derived in \cite{AFKdV}. Using the above three conditions, the real value of the classical mass can easily be shown by the following argument
    \begin{eqnarray}
         \int d^3 x \mathcal{H}[\mathcal{CPT}\phi(x)] &\overset{(1)}{=}& \int d^3 x \mathcal{H}^\dagger[\phi(-x)] =M_1^\dagger,\nonumber\\
        &\overset{(2)}{=}& \int d^3 x \mathcal{H}[\phi_2(-x)] =M_2,\nonumber\\
        \implies M_1^\dagger =M_2&\overset{(3)}{\implies} & M_1^\dagger =M_1,\nonumber
    \end{eqnarray}
    where numbers above the equal signs indicate the condition number. 
    
    The above analysis can be performed directly on the complex model. However, the non-Hermitian theory is only well-defined once the inner-product is identified. The modern way of the well-defined non-Hermitian quantum mechanics was first realised by Frederik Scholtz, Hendrik Geyer, and Fritz Hahne in 1992, \cite{Urubu}. The authors used the mathematical condition on the operator called the \textit{quasi-Hermiticity} (the term was first coined in \cite{Dieu}, but the metric was not given) to define the positive definite inner product. The quasi-Hermiticity is defined as a condition on the bounded linear operator of the Hilbert space $A: \mathcal{H}\rightarrow \mathcal{H}$, which satisfies 
\begin{itemize}
    \item[(i)] $\braket{v|\rho v}>0$ for all $\ket{v}\in\mathcal{H}$ and $\ket{v}\not= 0$.
    \item[(ii)] $\rho A = A ^\dagger \rho$ .
\end{itemize}
Where the bounded Hermitian linear operator $\rho: \mathcal{H}\rightarrow \mathcal{H}$ is often called the metric operator because the inner product is defined by the operator $\braket{\cdot | \cdot}_\rho:=\braket{\cdot|\rho \cdot}$ restores the Hermiticity of the operator. This result can be shown by using the condition(ii)
\small
\begin{eqnarray}
    \braket{v|Aw}_\rho \equiv \braket{v|\rho Aw} = \braket{v| A^\dagger \rho w} = \braket{Av| \rho w} = \braket{Av|w}_\rho , \nonumber
\end{eqnarray}
\normalsize
for all $\ket{v},\ket{w}\in \mathcal{H}$.
Note that the quasi-Hermiticity alone does not guarantee the real energy spectrum of the Hamiltonian. In fact, one requires two extra conditions. 
\begin{itemize}
    \item[(iii)] The metric operator is invertible.
    \item[(iv)] $\rho=\eta^\dagger \eta$.
\end{itemize}
The operator which satisfies only conditions (ii) and (iv) is referred to as the \textit{pseudo-Hermitian} operator, which was first introduced in \cite{pseudo1}. These extra conditions were considered in \cite{Urubu} to prove that, given a set of pseudo-Hermitian operators $\mathcal{A}=\{A_i\}$, the metric operator $\rho_\mathcal{A}$ which satisfies conditions (i), (ii), (iii) and (iv) for all operators of set $\mathcal{A}$ is uniquely determined if and only if all operators of the set $\mathcal{A}$ are irreducible on the Hilbert space $\mathcal{H}$. 
This procedure is analogous to the Dyson mapping first introduced by Freeman Dyson \cite{Dyson} used in the study of nuclear reaction \cite{Marumori1964,Beliaev1962,Janssen1971}, which maps the non-Hermitian operator $A$ to Hermitian operator $\eta^{-1} A \eta$ via \textit{Dyson map} $\eta$. The relation between the metric operator and the Dyson map is found by utilising the Hermiticity of the expression $\eta^{-1} A \eta$ in the following way
\small
\begin{equation}
    \eta^{-1} A \eta=(\eta^{-1} A \eta)^\dagger \implies A \eta^\dagger \eta = \eta^\dagger \eta A^\dagger \implies \eta^\dagger \eta = \rho.
\end{equation}
\normalsize
We will utilise this mapping to transform the non-Hermitian field-theoretic Hamiltonian to a Hermitian Hamiltonian. This procedure will resolve the issue of complex vacuum solution and Derrick's scaling argument, as we will see below. However, we note that the Dyson map used here introduces a negative kinetic sign in the kinetic term of one of the fields, indicating the ghost field problem. This issue is removed if one further diagonalise the Hamiltonian. Such diagonalisation can be realised via field-redefinition or via another Dyson map. A more detailed discussion of this is found in \cite{Fring2021}, and a Dyson map which diagonalise the free part of the non-Hermitian Hamiltonian is found in \cite{AlexandreSusy}.  
\section{Methods}\label{Section: Ch3 Sec1}
    In this section, we will summarise the method used in \cite{FringGoldstone,fring2020massive,fring2019pseudo,fring2020t} to find the masses of the Higgs particles, massive gauge particles and t'Hooft-Polyakov monopoles in non-Hermitian gauge field theory. We note that the explicit forms of the similarity transformation will not be discussed in this paper as non-Hermitian and Hermitian theories are isospectral as long as the $\mathcal{CPT}$ symmetry is preserved for Hamiltonian, Higgs particles and monopole solution. 
    
    We begin with the non-Hermitian local $SU(2)$ gauge theory with matter fields in the adjoint representation
    \begin{eqnarray}\label{Complex Lagrangian}
        \mathcal{L}_{2}^{\text{ad}} &=&\frac{1}{4}\text{Tr}\left( D\phi
        _{1}\right) ^{2}+\frac{%
        m_{1}^{2}}{4}\text{Tr}(\phi _{1}^{2})\label{Equation: Ch2 Sec5 adj_rep_action} \\
        &&-i\frac{\mu ^{2}}{2}\text{Tr}(\phi _{1}\phi _{2})-\frac{g}{64}\left[ \text{Tr}(\phi _{1}^{2})\right] ^{2}
        \notag\\
        &&+\frac{1}{4}\text{Tr}\left( D\phi _{2}\right) ^{2}+\frac{m_{2}^{2}}{4}\text{Tr}(\phi _{2}^{2})-\frac{1}{8}%
        \text{Tr}\left( F^{2}\right) .\nonumber
    \end{eqnarray}
    Here we take $g,\mu\in\mathbb{R}$, $m_i \in\mathbb{R}$ and discrete values $c_i \in \{-1,1\}$. The two fields $\{\phi_i \}_{i=1,2}$ are Hermitian matrices $\phi_i (t,\Vec{x}) \equiv \phi_i^a (t,\Vec{x}) T^a$, where $\phi_i^a (t,\Vec{x})$ is a real-valued field. The three generators $\{T^a\}_{a=1,2,3}$ of $SU(2)$ in the adjoint representation are defined by three Hermitian matrices of the form $(T^a)_{bc} = -i \epsilon_{abc}$, satisfying the commutation relation $[T^a, T^b]=i\epsilon^{abc} T^c$. One can check that $Tr(T^a T^b)=2\delta ^{ab}$. The field strength tensor is defined as $F_{\mu\nu} = \partial_\mu A_\nu - \partial_\nu A_\mu -i e [A_\mu,A_\nu]$, where the gauge fields are $A_\mu = A_\mu^a T^a$. The partial derivative is replaced with the covariant derivative $(D_{\mu }\phi _{i})^{a}:=\partial _{\mu }\phi _{i}^{a}+e\varepsilon
    _{abc}A_{\mu }^{b}\phi _{i}^{c}$ to compensate for the local symmetry group $SU(2)$.
        
    This action is invariant under the local $SU(2)$ transformation of the matter fields $\phi_i \rightarrow e^{i \alpha^a(x) T^a } \phi_i e^{-i \alpha^a(x) T^a }$ and gauge fields $A_\mu \rightarrow  e^{i \alpha^a(x) T^a }A_\mu e^{-i \alpha^a(x) T^a }+\frac{1}{e}\partial_\mu \alpha^a(x) T^a  $. It is also symmetric under modified $\mathcal{CPT}$ symmetry, which transforms two fields, $\phi_1$ and $\phi_2$ as
    \begin{eqnarray}\label{Equation: Ch3 Sec1 CPT transforamtion}
        &\mathcal{CPT}&: \phi_1 (t,\Vec{x})\rightarrow  \phi_1 (-t,-\Vec{x})\nonumber\\
        &&: \phi_2 (t,\Vec{x})\rightarrow  -\phi_2 (-t,-\Vec{x})\nonumber\\
        &&:i \rightarrow -i .
    \end{eqnarray}
    The equations of motion for the fields $\phi_i$ and $A_\mu$ of the Lagrangian (\ref{Equation: Ch2 Sec5 adj_rep_action}) are 
    \begin{eqnarray}\label{Equation: Ch3 Sec1 equations of motion}
        &\left(D_\mu D^\mu \phi_i\right)^a + \frac{\delta V }{\delta \phi^a_i}=0 ,
        \\ &D_\nu F^{\nu\mu}_a-e \epsilon_{abc}\phi^b_1 (D^\mu \phi)^c +e \epsilon_{abc}\phi^b_2 (D^\mu \phi)^c =0, \nonumber
    \end{eqnarray}
    where repeated indices are summed over.
    We perform the similarity transformation of the complex Lagrangian (\ref{Equation: Ch2 Sec5 adj_rep_action}) by momentary resorting to a quantum theory where we assume an equal time commutation relation between the fields $\phi_i^a$ and their canonical momenta $\Pi_i^a = \partial_0 \phi_i^a$, satisfying the commutation relation $[\phi_i^a (t,\Vec{x}) , \Pi_j^b (t,\Vec{y})] = \delta(\Vec{x}-\Vec{y})\delta_{ij}\delta_{ab}$. Using this relation, we can transform the corresponding complex Hamiltonian of the Lagrangian (\ref{Equation: Ch2 Sec5 adj_rep_action}) by 
    \begin{eqnarray}\label{Equation: Ch3 Sec1 similarity transformation}
        &H\rightarrow e^{\eta_{\pm}}He^{-\eta_{\pm}},\\
        &\eta_{\pm}  = \prod_{a=1}^3 \exp\left(\pm\frac{\pi}{2}\int d^3 x \Pi^a_2 \phi^a_2 \right),\nonumber
    \end{eqnarray}
    where $H$ is the field-theoretic Hamiltonian of our model (\ref{Complex Lagrangian}), obtained via Legendre transformation.
    This non-uniqueness of the metric is analogous to the non-uniqueness of the metric and its connection to the observables in the quantum mechanical setting discussed in \cite{Urubu,Musumbu2007}.  
    The adjoint action of $\eta_\pm$ maps the complex action in the equation (\ref{Equation: Ch2 Sec5 adj_rep_action}) into the following real action
        \begin{eqnarray}\label{Equation: Ch3 Sec1 real adj rep action}
            \mathfrak{s} &=&\int d^4 x ~ \frac{1}{4}Tr\left(D\phi_1\right)^2 -\frac{1}{4}Tr\left(D\phi_2\right)^2 \\
            &&+c_1 \frac{m_1^2}{4} Tr(\phi_1^2) - c_2 \frac{m_2^2}{4} Tr(\phi_2^2)\nonumber\\
            &&-c_3\frac{\mu^2}{2} Tr(\phi_1 \phi_2)-\frac{g}{64}\left(Tr(\phi_1^2)\right)^2-\frac{1}{8}Tr(F^2)\nonumber\\
            &\equiv & \int d^4 x ~~ \frac{1}{4}Tr\left(D\phi_1\right)^2 -\frac{1}{4}Tr\left(D\phi_2\right)^2 \nonumber\\
            &&- V -\frac{1}{8}Tr(F^2).\nonumber
        \end{eqnarray}
        The parameter $c_3$ indicates the different similarity transformations by taking the values $\pm 1$ for $\eta_\pm$, respectively. 
        
        For convenience, let us rewrite the above real action in terms of each component of the fields $\phi_i^a$ as 
        \begin{eqnarray}
        \mathfrak{l}_{2}^{\text{ad}}&=&\frac{1}{2}(D_{\mu }\phi _{i})^{a}\mathcal{I}_{ij}(D^{\mu}\phi _{j})^{a}+\frac{1}{2}\phi _{i}^{a}H_{ij}\phi _{j}^{a}\label{Equation: Ch2 Sec5 adj_rep_action_real}\\
        &&-\frac{g}{16}\left( \phi_{i}^{a}E_{ij}\phi _{j}^{a}\right) ^{2}-\frac{1}{4}F_{\mu\nu}^a F^{a\mu\nu},  \nonumber
        \end{eqnarray}%
        where the matrices $H,\mathcal{I}$ and $E$ are defined as
        \begin{eqnarray}
        &H:=\left( 
        \begin{array}{cc}
        m_{1}^{2} & -\mu ^{2} \\ 
        -\mu ^{2} & -m_{2}^{2}%
        \end{array}%
        \right) ~,~\mathcal{I}:=\left( 
        \begin{array}{cc}
        1 & 0 \\ 
        0 & -1%
        \end{array}%
        \right), \\
        &E:=\left( 
        \begin{array}{cc}
        1 & 0 \\ 
        0 & 0%
        \end{array}%
        \right).\nonumber
    \end{eqnarray}
    \subsection{Higgs and gauge masses}
    Next, we define the trivial solution of the equations of motion by solving $\delta V = 0$ and $D_\mu \phi_i =0$. Such vacuum is often referred to as Higgs vacuum. The first equation can be simplified by choosing an Ansatz $(\phi_i^0)^a (t,\Vec{x})= h_i^0  \hat{r}^a (\Vec{x})$ where $\hat{r}=(x,y,z)/\sqrt{x^2 +y^2 +z^2}$ and $\{h^0_i\}$ are some constants to be determined. Note that the vacuum solution has a rotational symmetry $SO(3)$ since $\hat{r}^a \hat{r}^a = 1$. Inserting this Ansatz into the equation (\ref{Equation: Ch2 Sec5 adj_rep_action_real}), we find 
        \begin{eqnarray}
            V = -\frac{1}{2} h_i H_{ij}h_j +\frac{g}{16} h_1^4 .
        \end{eqnarray}
        Then the vacuum equation $\delta V =0$ is reduced to simple coupled third order algebraic equations 
        \begin{eqnarray}
            \frac{g}{4} (h_1^0)^3 - c_1 m_1^2 h_1^0 + c_3 \mu^2 h_2^0 &=&0, \nonumber\\
            c_2 m_2^2 h_2^0 +c_3 \mu^2 h_1^0&=&0,\label{Equation: Ch3 Sec1 higgs vacuum 2} \\
            D_\mu \phi_\alpha &=&0.
        \end{eqnarray}
        The resulting vacuum solutions are
        \begin{eqnarray}
            &\!\!\!\!h_2^0 = -\frac{c_2 c_3 \mu^2}{m_2^2}h_1^0 ,~ (h_1^0)^2 =4\frac{c_2 \mu^4 + c_1 m_1^2 m_2^2}{gm_2^2}:= R^2,~~~~ \label{Equation: Ch3 Sec1 higgs vacuum}\\
            &(A_i^0)^a  = -\frac{1}{er}\epsilon^{iaj}\hat{r}^j+ \hat{r}^a \mathcal{A}_i,~(A_0^0)^a  =0.\nonumber 
        \end{eqnarray}
        The $\mathcal{A}_i$ are arbitrary functions of space-time. The Higgs particle can be identified with the fundamental fields of the theory after spontaneous symmetry breaking of the continuous symmetry $SU(2)$ by Taylor expanding around the vacuum solution. Performing a Taylor expansion around the Higgs vacuum and keeping a focus on the second-order terms with only matter fields $\phi_i^a$, the real Lagrangian (\ref{Equation: Ch2 Sec5 adj_rep_action_real}) contains the following term 
        \begin{eqnarray}
            \mathfrak{s} &=& \int d^4 x~\frac{1}{2}  \phi_i^a \left(-\partial_\mu\partial^\mu I_{ij}\delta^{ab}-\Tilde{H}_{ij}^{ab}\right) \phi_j^b+\dots ,\nonumber
        \end{eqnarray}
        where $\Tilde{H}$ is a $6\times 6$ block diagonal Hermitian matrix. After diagonalising the above term by redefining the fields with the eigenvectors of the non-Hermitian mass matrix $M_{ij}^{ab}:= I_{ik}\delta^{ac}\Tilde{H}_{kj}^{cb}$, we find that the masses of the fundamental fields after the symmetry breaking to be equal to the eigenvalues of $M_{ij}^{ab}$ given as
        \begin{equation}\label{higgs mass}
            m_0^2 = c_2 \frac{\mu^4 - m_2^4}{m_2^2} ~,~~~ m_{\pm}^2 = K \pm \sqrt{K^2 + 2L},
        \end{equation}
        where $K= c_1 m_1^2 -c_2 \frac{m_2^2}{2} +\frac{3 \mu^4}{2 c_2 m_2^2}$ and $L=\mu^4 + c_1 c_2m_1^2 m_2^2$. Notice that we only find three non-zero eigenvalues. The redefined fields with zero masses (eigenvalues) are called Goldstone fields, which can be absorbed into gauge fields $A_\mu^a$ by defining the new massive gauge fields. This process of giving mass to the previously massless fields is called the Higgs mechanism. The mass of the gauge fields can be found by expanding the kinetic term of $\phi$ around the Higgs vacuum $\phi^0_i = h_i^0 \hat{r}^a$. Without loss of generality, we can choose a particular direction of the vacuum by taking $\hat{r}=(0,0,1)^T$. This is possible due to the symmetry of the $SO(3)$ vacuum as discussed above. Keeping the term only quadratic in the gauge field, we find 
        \begin{eqnarray}
            &&\frac{1}{2}(D_\mu \phi_i + D_\mu \phi_i^0)^a\mathcal{I}_{ij}(D_\mu \phi_i + D_\mu \phi_i^0)^a  \\ &=&\frac{1}{2}\left(eA_\mu \times \phi^0_i\right)^a\mathcal{I}_{ij}\left(eA_\mu \times \phi^0_j\right)^a +\dots\nonumber\\
            &=& \frac{1}{2}e^2 h_i^0 \mathcal{I}_{ij}h^0_j \left(A^1_\mu {A^1}^\mu +A^2_\mu {A^2}^\mu  \right)+\dots \nonumber\\
            &=& \frac{1}{2}m_g^2 \left(A^1_\mu {A^1}^\mu +A^2_\mu {A^2}^\mu  \right)+\dots ,\nonumber
        \end{eqnarray}
        where the mass of the gauge field is identified to be $m_g :=\sqrt{ h_i^0 \mathcal{I}_{ij}h^0_j} =e\frac{R \sqrt{m_2^4 - \mu^4}}{m_2^2}$.
    \subsection{t'Hooft-Polyakov Monopole}
        To find the monopole solutions, let us consider the following Ansatz
        \begin{eqnarray}\label{Equation: Ch3 Sec1 Spherical Ansatz}
            &(\phi^{cl}_i)^a (\Vec{x})=  h_i(r) \hat{r}^a ~,~ (A_i^{cl})^a = \epsilon^{iaj}\hat{r}^j A(r)~,\\ &(A_0^{cl})^a =0,\nonumber
        \end{eqnarray}
        where the subscript $cl$ denotes the classical solutions to the equations of motion (\ref{Equation: Ch3 Sec1 equations of motion}). The difference between this Ansatz (\ref{Equation: Ch3 Sec1 Spherical Ansatz}) and the Higgs vacuum (\ref{Equation: Ch3 Sec1 higgs vacuum}) is that the quantity $h_i$ now depends on the spatial radius $h_i = h_i(r)$. 
        Here we are only considering the static Ansatz to simplify our calculation, but one may, of course, also consider the time-dependent solution by utilising the Lorentz symmetry of the model and performing a Lorentz boost. According to Derrick's scaling argument \cite{derrick1964comments}, for the monopole solution to have finite energy, we require the two matter fields of the equation (\ref{Equation: Ch3 Sec1 Spherical Ansatz}) to approach the vacuum solutions in the equation (\ref{Equation: Ch3 Sec1 higgs vacuum}) at spatial infinity
        \begin{eqnarray}\label{Equation: Ch3 Sec1 asymptotic condition}
            \lim_{r\rightarrow\infty} h_1 (r) =h_1^{0\pm}= \pm R~,~\\ \lim_{r\rightarrow\infty} h_2 (r) =h_2^{0\pm}= \mp \frac{c_2 c_3\mu^2}{m_2^2} R. \nonumber
        \end{eqnarray}
        Also, notice that at some fixed value of the radius $r$, the vacuum solutions $\phi_\alpha^0$ and monopole solutions $\phi_\alpha^{cl}$ both belongs to the 2-sphere in the field configuration space. For example, $\phi_1^0$ belongs to the 2-sphere with radius $R$ because $(\phi_1^0)^2 = R^2$. Therefore, solutions $\phi^{cl}_i$ can be seen as a mapping between 2-sphere in space-time (where the radius is given by the profile function $h_i$) to 2-sphere in field configuration space. Such mapping has a topological number called the winding number $n\in\mathbb{Z}$, which can be explicitly realised by redefining the unit vector $\hat{r}^a$ as
        \begin{equation}\label{Equation: Ch3 Sec 2 radial unit vector}
            \hat{r}_n^a = \left(\begin{array}{c}
                \sin(\theta)\cos(n \varphi)  \\
                \sin(\theta)\sin(n \varphi) \\
                \cos(\theta)
            \end{array}\right).
        \end{equation}
        Therefore different $n$ represent topologically inequivalent solutions. 
        
        Since we require the monopole and vacuum solutions to smoothly deformed into each other at spacial infinity, both solutions need to share the same winding number. It is important to note that winding numbers of $\phi_1 $ and $\phi_2$ need to be equal to satisfy $D\phi_1 = D\phi_2 =0$, and therefore we will denote the winding numbers of $\phi_1$ and $\phi_2$ as $n$ collectively. If they are not equal, we would have $D\phi_1 =0$ but $D\phi_2 \not=0$. Next, let us insert our Ansatz equation (\ref{Equation: Ch3 Sec1 Spherical Ansatz}) into the equations of motion equation (\ref{Equation: Ch3 Sec1 equations of motion}). We will also redefine the Ansatz for the gauge fields to be $A^a_i = \epsilon^{aib}\hat{r}^b \left(\frac{1-u(r)}{e r}\right),~A_0^a =0$, which are more in line with the original Ansatz given in \cite{prasad1975exact,bogomol1976stability}, compared to  equation (\ref{Equation: Ch3 Sec1 Spherical Ansatz}).
        Inserting these expressions into the equations of motion equation (\ref{Equation: Ch3 Sec1 equations of motion}), we find 
        \begin{eqnarray}
            u^{''}(r)+\frac{u(r)\left[1-u^2 (r)\right]}{r^2}\label{Equation: Ch3 Sec1 t'Hooft Polyakov differential equations 1}\\
            +\frac{ e^2 u(r)}{2} \left\{h_2^2 (r) - h_1^2 (r)\right\} =0,\nonumber
        \end{eqnarray}
        \begin{eqnarray}
            &h_1^{''}(r)+\frac{2h_1^{'} (r)}{r}-\frac{2 h_1(r)u^2(r)}{r^2}\label{Equation: Ch3 Sec1 t'Hooft Polyakov differential equations 2}\\
            &+g\left\{- c_1 \frac{m_1^2}{g} h_1 (r)+c_3 \frac{  \mu^2}{g} h_2 (r)+ \frac{1}{4}h_1^3  (r)\right\}=0,\nonumber
        \end{eqnarray}
        \begin{eqnarray}
            h_2^{''}(r)+\frac{2h_2^{'} (r)}{r}-\frac{2 h_2(r)u^2(r)}{r^2}\label{Equation: Ch3 Sec1 t'Hooft Polyakov differential equations 3} \\
            +c_2 m_2^2 \left\{ h_2 (r)+ c_3 \frac{\mu^2}{m_2^2} h_1 (r)\right\}=0 .\nonumber
        \end{eqnarray}
        Notice that these differential equations are similar to the ones discussed in \cite{prasad1975exact,bogomol1976stability}, but with the extra field $h_2$ and extra differential equation (\ref{Equation: Ch3 Sec1 t'Hooft Polyakov differential equations 3}). In the Hermitian model, the exact solutions to the differential equations were found by taking the parameter limit called the BPS limit \cite{prasad1975exact,bogomol1976stability}, where parameters in the Hermitian model are taken to zero while keeping the vacuum solution finite. Here we will follow the same procedure and take the parameter limit where quantities in the curly brackets of equations (\ref{Equation: Ch3 Sec1 t'Hooft Polyakov differential equations 2}) and (\ref{Equation: Ch3 Sec1 t'Hooft Polyakov differential equations 3}) vanish but keep the vacuum solutions equation (\ref{Equation: Ch3 Sec1 higgs vacuum}) finite. We will see in section \ref{BPS limit section} that we also find the approximate solutions in this limit.
     \subsection{The energy bound}\label{section:energy_bound}
        Surprisingly, by utilising Derrick's scaling argument, one can find the lower bound of the monopole energy without the explicit form of the solution. 
        
        The energy of the monopole can be found by inserting the monopole solution into the corresponding Hamiltonian of equation (\ref{Equation: Ch3 Sec1 real adj rep action}).
        \begin{eqnarray}
            \!\!\!\!\mathfrak{h} &=& \int d^3 x ~~Tr\left(E^2\right) + Tr\left(B^2\right) \\
            &&+Tr\left\{(D_0 \phi_1)^2\right\}+Tr\left\{(D_i \phi_1)^2\right\}\nonumber\\
            &&-Tr\left\{(D_0 \phi_2)^2\right\}-Tr\left\{(D_i \phi_2)^2\right\}+V,\nonumber
        \end{eqnarray}
        where $E,B$ are ${E^i}_a ={F_a}^{0i}$ , ${B^i}_a =- \frac{1}{2}\epsilon^{ijk}F_a^{ jk}$, $i,j,k \in \{1,2,3\}$. The gauge is fixed to be the radiation gauge (i.e ${A_a}^0 =0 , \partial_i {A_a}^i =0$). Notice that our monopole Ansatz equation (\ref{Equation: Ch3 Sec1 higgs vacuum 2}) is static with no electric charge $E_i^a=0$ and therefore, the above Hamiltonian reduces to 
        \begin{eqnarray}
            E &=& \int d^3 x  ~~ Tr\left({B}^2\right) +Tr\left\{(D_i \phi_1)^2\right\}\\
            &&-Tr\left\{(D_i \phi_2)^2\right\}+V\nonumber\\
            &=& 2\int d^3 x  ~~ {B_i}^a {B_i}^a +(D_i \phi_1)^a (D_i \phi_1)^a \nonumber\\
            &&-(D_i \phi_2)^a (D_i \phi_2)^a +\frac{1}{2}V.\nonumber
        \end{eqnarray}
        Here, we simplified our expression by dropping the superscripts $A^{cl}_i \rightarrow A_i$ , $\phi_\alpha^{cl} \rightarrow \phi_\alpha$. We also keep in mind that these fields depend on the winding numbers $n\in \mathbb{Z}$. In the Hermitian model (i.e. when $\phi_2 =0$), one can rewrite the kinetic term as $B^2 + D\phi^2 = (B-D\phi)^2 + 2 BD\phi$ and find the lower bound to be $\int 2 BD\phi$. Here we will follow a similar procedure but introduce some arbitrary constant $\alpha,\beta \in \mathbb{R}$ such that $B^2 = \alpha^2 B -\beta^2 B$ where $\alpha^2 -\beta^2 =1$. This will allow us to rewrite the above energy as
        \begin{eqnarray}\label{energy of the monopole}
            E &=& 2\int d^3 x ~~\alpha^2 \left\{{B_i}^a + \frac{1}{\alpha}(D_i \phi_1 )^a\right\}^2 \\
            &&-\beta^2 \left\{{B_i}^a + \frac{1}{\beta}(D_i \phi_2 )^a\right\}^2 \nonumber\\ 
            &&+ 2\left\{-\alpha {B_i}^a (D_i \phi_1 )^a + \beta  {B_i}^a (D_i \phi_2 )^a \right\}+\frac{1}{2}V.\nonumber
        \end{eqnarray}
        To proceed from here, we need to assume extra constraints on $\alpha$ and $\beta$ such that the following inequalities are true  
        \begin{eqnarray}
            \int d^3 x ~\alpha^2 \left\{{B_i}^a + \frac{1}{\alpha}(D_i \phi_1 )^a\right\}^2 &&\label{boundary condition 1}\\
            -\beta^2 \left\{{B_i}^a + \frac{1}{\beta}(D_i \phi_2 )^a\right\}^2&\geq &0,\nonumber \\
            \int d^3 x V &\geq& 0 . \label{boundary condition 2}
        \end{eqnarray}
        With these constraints we can now write down the lower bound of the monopole as 
        \small
        \begin{eqnarray}
            E &\geq &2\int d^3 x  ~ \left\{-\alpha {B_i}^a (D_i \phi_1 )^a + \beta {B_i}^a (D_i \phi_2 )^a \right\}\\
            &=&2\int d^3 x~ - \alpha \left\{{B_i}^a \partial_i \phi_1^a + e {B_i}^a \epsilon^{abc}{A_i}^b \phi_1^c\right\}\nonumber\\
            &&+ \beta \left\{{B_i}^a \partial_i \phi_2^a + e {B_i}^a \epsilon^{abc}{A_i}^b \phi_2^c\right\}\nonumber\\
            &=&2\int d^3 x~  -\alpha \left\{{B_i}^a \partial_i \phi_1^a + \left(-e\epsilon^{abc}{A_i}^b {B_i}^c\right)\phi_1^a\right\}\nonumber\\
            &&+ \beta \left\{{B_i}^a \partial_i \phi_2^a +\left(-e\epsilon^{abc}{A_i}^b {B_i}^c\right)\phi_1^a \phi_2^c\right\}\nonumber\\
            &=&2\int d^3 x~ -\alpha \left\{{B_i}^a \partial_i \phi_1^a +\partial_i {B_i}^a \phi_1^a\right\}\nonumber\\
            &&+\beta \left\{{B_i}^a \partial_i \phi_2^a + \partial_i {B_i}^a \phi_1^a\right\}\nonumber\\
            &=& 2\int d^3 x ~ -\alpha \partial_i \left({B_i}^a {\phi_1}^a\right)+\beta \partial_i \left({B_i}^a {\phi_2}^a\right)\nonumber\\
            &=&\lim_{r\rightarrow \infty}\left(-2\alpha \int_{S_r} dS_i {B_i}^a {\phi_1}^a + 2\beta \int_{S_r} dS_i {B_i}^a {\phi_2}^a\right),\nonumber
        \end{eqnarray}
        \normalsize
        where in the fourth line, we used $D_i B_i^a =0 $, which can be shown from the Bianchi identity $D_\mu \epsilon^{\mu\nu\rho\sigma}F^a_{\rho\sigma}=0$. The last line is obtained by using the Gauss theorem at some fixed value of the radius $r$. Since the $\phi_i^a$ in the integrand is only defined over the 2-sphere with a large radius, we can use the asymptotic conditions (\ref{Equation: Ch3 Sec1 asymptotic condition}) and replace the monopole solutions $\{\phi_\alpha^a, B_i^a\}$ with the Higgs vacuum $\{(\phi_\alpha^0)^a ,(B_i^0)^a\}$
        \small
        \begin{eqnarray}\label{energy bound}
            E&\geq & \left(-2\alpha {\phi^0_1}^a  + 2\beta {\phi^0_2}^a\right)\lim_{r\rightarrow\infty}\int_{S_r} dS_i (B_i^0)^a \\
            &=& \left(\mp2\alpha R \hat{r}^a_n \mp 2 \beta \frac{c_2 c_3\mu^2}{m_2^2} R \hat{r}^a_n\right) \lim_{r\rightarrow\infty}\int_{S_r} dS_i (B_i^0)^a ,\nonumber
        \end{eqnarray}
        \normalsize
        where the upper and lower signs of the above energy correspond to the upper and lower signs of the vacuum solutions in equation (\ref{Equation: Ch3 Sec1 higgs vacuum}). The explicit value of $B_{0i}^a$ can be obtained by inserting the Higgs vacuum (\ref{Equation: Ch3 Sec1 higgs vacuum}) into the definition of the magnetic field 
        \begin{equation}\label{Equation: Ch3 Sec1 definition of B 2}
            B_{i}^a = - \frac{1}{2} {\epsilon_{i}}^{jk} \left(\partial_j A_k -\partial_k A_j + e A_j\times A_k \right)^a .
        \end{equation}
        After a lengthy calculation, this expression can be simplified to $B_{0i}^a = \hat{\phi^0}^{a} b_i = \hat{r}^a_{n} b_i$, where $\hat{\phi^0}^{a}$ is a normalised solution $\sum_a \hat{\phi^0}^{a} \hat{\phi^0}^{a} =1$. The $b_i$ is defined as 
        \begin{equation}\label{Equation: Ch3 Sec1 definition of B}
            b_i \equiv -\frac{1}{2} \epsilon_{ijk} \left\{\partial^j \mathcal{A}^k -\partial^k \mathcal{A}^j +\frac{1}{e} \hat{r}_n\cdot \left(\partial^j \hat{r}_n \times \partial^k \hat{r}_n\right) \right\} .
        \end{equation}
        Where $\mathcal{A}$ was defined in equation (\ref{Equation: Ch3 Sec1 higgs vacuum}).
        Notice that integrating the first term over the 2-sphere gives zero by Stoke's theorem  $\int_S \partial \times \mathcal{A} = \int_{\partial S} \mathcal{A} =0$, where one can show that Stoke's theorem on the closed surface gives zero by dividing the sphere into two open surfaces. The second term is a topological term which can be evaluated as
        \begin{equation}\label{Equation: Ch3 Sec2 winding number}
            \int dS_i B_i = -\frac{4 \pi n}{e} .
        \end{equation}
        The explicit calculation is in \cite{arafune1975topology}.
        This is the magnetic charge of the monopole solutions. Therefore integer $n$, which corresponds to the winding number of the solution, comes from the Ansatz $B_i ^a = \hat{\phi^0}^{a} b_i $. In our case, there is an ambiguity of whether to choose $B_i ^a = \hat{\phi_1^0}^{a} b_i $ or $B_i ^a = \hat{\phi_2^0}^{a} b_i $. Now we see explicitly the reason why we choose to keep the same integer values for solutions $\phi_1^0$ and $\phi^0_2$. If the integer values of $\hat{r}_n^a$ in solutions $\phi_1^0$, $\phi^0_2$ are different, then the integration $\int_{S_r} dS_i (B_i^0)^a$ will be different, leading to inconsistent energy. 
        
        Finally, we find our lower bound of the monopole energy 
        \begin{eqnarray}\label{energy bound of the monopole}
            E &\geq & \mp2 R \left(\alpha +\beta \frac{c_2 c_3 \mu^2}{m_2^2}\right)\hat{r}^a_n \hat{r}^a_{n} \left(\frac{-4\pi n }{e}\right)\\
            &=& \frac{\pm 8\pi n R }{e} \left(\alpha+\beta \frac{c_2 c_3\mu^2}{m_2^2}\right)\nonumber .
        \end{eqnarray}
        Notice that we have some freedom to choose $\alpha ,\beta\in\mathbb{R}$ as long as our initial assumptions (\ref{boundary condition 1}) are satisfied. We will see in the next section that we can take a parameter limit of our model, which saturates the above inequality and gives exact values to $\alpha$ and $\beta$. 
        \subsection{The fourfold BPS scaling limit}\label{BPS limit section}
    Our main goal is now to solve the coupled differential equations (\ref{Equation: Ch3 Sec1 t'Hooft Polyakov differential equations 1})-(\ref{Equation: Ch3 Sec1 t'Hooft Polyakov differential equations 3}). Prasad, Sommerfield, and Bogomolny \cite{prasad1975exact,bogomol1976stability} managed to find the exact solution by taking the parameter limit, which simplifies the differential equations. The multiple scaling limit is taken so that all the parameters of the model tend to zero with some combinations of the parameter remaining finite. The combinations are taken so that the vacuum solutions stay finite in this limit. Inspired by this, we will take here a fourfold scaling limit
    \begin{equation}\label{modified BPS limit}
        g,m_1,m_2,\mu\rightarrow 0 ~,~~~ \frac{m_1^2}{g}<\infty ~,~~~ \frac{\mu^2}{g}<\infty ~,~~~ \frac{\mu^2}{m_2^2}<\infty .
    \end{equation}
    This will ensure that the vacuum solutions equation (\ref{Equation: Ch3 Sec1 higgs vacuum}) stays finite, but crucially the curly bracket parts in equations (\ref{Equation: Ch3 Sec1 t'Hooft Polyakov differential equations 2}) and (\ref{Equation: Ch3 Sec1 t'Hooft Polyakov differential equations 3}) vanish. There is a physical motivation for this limit in which the mass ratio of the Higgs and gauge mass are taken to be zero (i.e. $m_{\text{Higgs}}<< m_g$) as described in \cite{kirkman1981asymptotic}. We will see in the next section that the same type of behaviour is present in our model, hence justifying equation (\ref{modified BPS limit}).  The resulting set of differential equations, after taking the BPS limit, is similar to the ones considered in \cite{prasad1975exact,bogomol1976stability} with the slightly different quadratic term in equation (\ref{Equation: Ch3 Sec1 t'Hooft Polyakov differential equations 1}). It is natural to consider a similar Ansatz as given in \cite{prasad1975exact,bogomol1976stability} 
    \begin{eqnarray}
        u(r) &=& \frac{evr}{\sinh{(evr)}},\label{BPS Ansatz 1}\\
        h_1 (r)&=&   -\alpha f(r),\label{BPS Ansatz 2}\\
        h_2 (r)&=& -\beta f(r),\label{BPS Ansatz 3}
    \end{eqnarray}
   where $\alpha,\beta \in \mathbb{R}$ were introduced in section \ref{section:energy_bound} and $f(r)\equiv \left\{v\coth\left(evr\right)-\frac{1}{er}\right\}$. One can check that this Ansatz indeed satisfies differential equations equation (\ref{Equation: Ch3 Sec1 t'Hooft Polyakov differential equations 1})-(\ref{Equation: Ch3 Sec1 t'Hooft Polyakov differential equations 3}) in the BPS limit. We have decided to put a prefactor $\alpha$ and $\beta$ in front of equations (\ref{BPS Ansatz 2}) and (\ref{BPS Ansatz 3}) to satisfy the differential equation (\ref{Equation: Ch3 Sec1 t'Hooft Polyakov differential equations 1}). Note that if we take $\alpha=1$, we get exactly the same as given in \cite{prasad1975exact,bogomol1976stability}, which is known to satisfy the first-order differential equation called the Bogomolny equation $B_i -D_i \phi =0$. The Ansatz (\ref{BPS Ansatz 1})-(\ref{BPS Ansatz 3}) only differs from the ones given in \cite{prasad1975exact,bogomol1976stability} by the prefactors $\alpha$ and $\beta$, and therefore our Ansatz should satisfy the Bogomolny equation with the appropriate scaling to cancel the prefactor in equations (\ref{BPS Ansatz 2}) and (\ref{BPS Ansatz 3})    
    \begin{eqnarray}
        B_i^b + \frac{1}{ \alpha} (D_i \phi_1)^b &=&0,\\
        B_i^b + \frac{1}{\beta} (D_i \phi_2)^b&=&0,
    \end{eqnarray}
    where $\phi_\alpha \equiv h_\alpha (r) \hat{r}_{n}$. 
    If we compare these equations to the terms appearing in the energy of the monopole equation (\ref{energy of the monopole}), then we can saturate the inequality in equation (\ref{energy bound of the monopole}) by
    \begin{eqnarray}\label{saturated_energy}
        E[\phi_1 ,\phi_2]=\frac{\pm8\pi n R}{e} \left(\alpha+\beta \frac{c_2 c_3\mu^2}{m_2^2}\right),
    \end{eqnarray}
    where upper and lower signs correspond to the vacuum solutions equation (\ref{Equation: Ch3 Sec1 higgs vacuum}), when taking the square root. We can calculate the explicit forms of $\alpha$ and $\beta$ by comparing the asymptotic conditions in equation (\ref{Equation: Ch3 Sec1 asymptotic condition})
    \begin{eqnarray}\label{matching_asymptotic_condition_1}
        \lim_{r\rightarrow \infty}h_1^\pm &=&h_1^{0\pm}= \pm R ,\\ \lim_{r\rightarrow \infty}h_2^\pm &=&h_2^{0\pm}= \mp \frac{c_2 c_3\mu^2}{m_2^2}R,\nonumber
    \end{eqnarray}
    with the asymptotic values of equations (\ref{BPS Ansatz 1})-(\ref{BPS Ansatz 3})
    \begin{eqnarray}\label{matching_asymptotic_condition_2}
        &\lim_{r\rightarrow \infty} u(r) =0 ~,~ \lim_{r\rightarrow \infty} h_1^\pm(r) =-\alpha v ,\\ &\lim_{r\rightarrow \infty} h_2^\pm(r) =-\beta v.\nonumber
    \end{eqnarray}
    By Derrick's scaling argument, the two asymptotic 
   values (\ref{matching_asymptotic_condition_1}) and (\ref{matching_asymptotic_condition_2}) should match, resulting in algebraic equations for $\alpha$ and $\beta$. Using $\alpha^2 - \beta^2 =1$ and assuming $m_2^4 \geq \mu^4$, we find the four set of real solutions 
    \begin{equation}\label{explict values of a and v}
        \alpha = \mp(\pm)\frac{m_2^2}{l}~,~~~ v=(\pm) \frac{R l}{m_2^2}~,~~~\beta =\pm (\pm)\frac{c_2 c_3 \mu^2}{l} ,
    \end{equation}
    where $l=\sqrt{m_2^4 -\mu^4}$. The plus-minus signs in the brackets correspond to the two possible solutions to the algebraic equation $\alpha^2 -\beta^2 =1$. These need to be distinguished from the upper and lower signs of $\alpha$ and $\beta$, which correspond to the vacuums solutions (\ref{Equation: Ch3 Sec1 higgs vacuum}). Inserting the explicit values of $\alpha$ and $\beta$ to the energy equation (\ref{saturated_energy}) we find
    \begin{eqnarray}\label{Equation: Ch3 Sec1 all_possible_energies}
        E [\phi_1 ,\phi_2]&\equiv & (\pm)\frac{8\pi n R}{ e m_2^2} \left(\frac{ -m_2^4 + \mu^4}{l}\right)\\
        &=&(\pm)\frac{-8\pi n R}{ e m_2^2} l,\nonumber
    \end{eqnarray}
    with corresponding solutions 
    \begin{eqnarray}
        h_1^\pm (r) &=& \pm (\pm)\frac{m_2^2}{l}\left[\frac{R l}{m_2^2}\text{coth}\left(\frac{eRl}{m_2^2}r\right)-\frac{1}{er }\right],\label{Equation: Ch3 Sec1 monopole solutions form 1}\\
        h_2^\pm (r) &=& \mp(\pm)\frac{c_2 c_3 \mu^2}{l}\left[\frac{R l}{m_2^2}\text{coth}\left(\frac{eRl}{m_2^2}r\right)-\frac{1}{er }\right].\nonumber
    \end{eqnarray}
    It is crucial to note that although it seems like there are two monopole solutions $\{h_1^\pm, h_2^\pm\}$, the two solutions are related non-trivially in their asymptotic limit by the constraint $\lim_{r\rightarrow \infty} h_2^\pm =( -c_2 c_3 \mu^2 / m_2^2)\lim_{r\rightarrow \infty} h_1^\pm$ given in equation (\ref{Equation: Ch3 Sec1 higgs vacuum 2}). For example, one can not choose $\{h_1^+ , h_2^-\}$ as a solution as this will break the asymptotic constraint. 
    
    The solution (\ref{Equation: Ch3 Sec1 monopole solutions form 1}) can be constrained further by imposing that the energy (\ref{Equation: Ch3 Sec1 all_possible_energies}) is real and positive. 
    \small
    \begin{eqnarray}
        E[\phi_1 ,\phi_2]>0 \implies -(\pm)\frac{8\pi n R}{ e m_2^2} l \implies - (\pm)n >0.\label{Equation: Ch3 Sec1 positve monopole mass requirement}
    \end{eqnarray}
    \normalsize
    Therefore we can ensure positive energy if $(\pm) = \text{sign}(n)$. The final form of the monopole solution with positive energy are 
    \small
    \begin{eqnarray}
        h_1^\pm (r) &=& \pm \text{sign}(n)\frac{m_2^2}{l}\left[\frac{R l}{m_2^2}\text{coth}\left(\frac{eRl}{m_2^2}r\right)-\frac{1}{er }\right],\label{Equation: Ch3 Sec1 monopole solutions}\\
        h_2^\pm (r) &=& \mp\text{sign}(n)\frac{c_2 c_3 \mu^2}{l}\left[\frac{R l}{m_2^2}\text{coth}\left(\frac{eRl}{m_2^2}r\right)-\frac{1}{er }\right].\nonumber
    \end{eqnarray}
    \normalsize
    with energy $E = 8|n|\pi l R/e m_2^2 $. We conclude this subsection by observing that the above solution depends on the parameter $c_3$, which takes value $\{-1,1\}$ depending on the choice of the similarity transformation. Choosing a different values of $c_3$ also result in a different asymptotic values (\ref{matching_asymptotic_condition_1}), meaning solutions for $c_3=1$ and $c_3=-1$ are topologically different. Since the energy is independent of $c_3$, two distinct solutions share the same energy. Respecting one of the main features of similarity transformation, which is to preserve the energy of the transformed Hamiltonian. 
    
    In the next section, we will investigate in detail how the solution changes and a new $\mathcal{CPT}$ symmetry emerges by changing the parameter values.  
    
    \section{Results and Discussion}\label{Section: Ch3 Sec1 Subsec5}
    
    This section will investigate the behaviour of solution (\ref{Equation: Ch3 Sec1 monopole solutions}) in different regimes of the parameter spaces. We will compare the physical regions of gauge particles, Higgs particles and monopoles found in the previous section. We will see that the two regions coincide, but the solutions in different regions possess different $\mathcal{CPT}$ symmetries. Different symmetries of solutions in different regions are not coincident, but the consequence of the three reality conditions stated in the introduction. In fact, it is deeply related to the real value of energy, which will be discussed extensively in \cite{FringSkyrmion}.

    \subsection{Higgs mass and exceptional points}\label{Section: Higgs recap}
    
        Let us recall the masses of the particles and monopole
        \begin{eqnarray}
            &m_0^2 = c_2 \frac{\mu^4 - m_2^4}{m_2^2} ~,~ m_{\pm}^2 = K \pm \sqrt{K^2 + 2L},\\
           & m_g = e\frac{R l}{m_2^2}, M_{\text{mono}}=\frac{8|n|\pi l R}{em_2^2}.\nonumber
        \end{eqnarray}
        where $K= c_1 m_1^2 -c_2 \frac{m_2^2}{2} +\frac{3 \mu^4}{2 c_2 m_2^2}$ and $L=\mu^4 + c_1 c_2m_1^2 m_2^2$. Notice that the masses do not depend on $c_3$, meaning they do not depend on the similarity transformation as expected. We also comment that in the BPS limit, we have $m_0 = m_\pm =0$, but $m_g$ and $M_\pm$ stays finite, such that the ratios $m_{\text{Higgs}}/m_g$ vanish in the BPS limit. This is in line with the Hermitian case \cite{kirkman1981asymptotic}, providing the physical interpretation with $m_{Higgs}<<m_g$ for the BPS limit.
        
        One may notice that when $c_2=1$, requiring positive mass $m_0^2 >0 $, implies that $\mu^4 - m_2^4 >0$. This means the quantity $l=\sqrt{m_2^4 - \mu^4}$ is purely imaginary. One may then discard this region as unphysical. However, we will see in the next section that there is a disconnected region beyond $\mu^4 - m_2^4 >0$, which admit real energy because $R$ also becomes purely complex. This is not coincident, and in fact, we will see an emerging new $\mathcal{CPT}$ symmetry for the monopoles.
        
\begin{figure*}[ht]
                \centering                \includegraphics[scale=0.6]{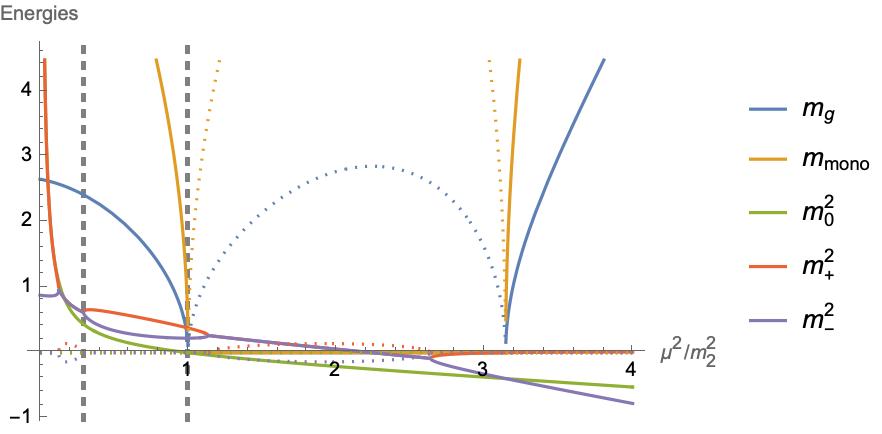}
                \caption{Monopole, gauge and Higgs masses plotted for $m_1^2/g=-0.44, \mu/g=-0.14, e=2, c_1=-c_2=-1$. The solid line represents the real part, and the dotted line represents the imaginary part of the masses. The dotted vertical lines indicate the boundaries of the physical regions where all the masses acquire real positive values.}
                \label{Figure: Ch3 Sec1 monopole vs gauge vs Higgs}
            \end{figure*}

        In the rest of the section, we will exclusively focus on the monopole and gauge masses. The main message of this section is the emerging symmetry responsible for the real value of the monopole masses. The requirement to make the whole theory physical demands also to consider the intersection of the physical regions between monopole masses and Higgs masses. As an example, we plot all the masses of the theory in figure \ref{Figure: Ch3 Sec1 monopole vs gauge vs Higgs}. 
            \begin{figure*}[ht]
                \centering
                \includegraphics[scale=0.47]{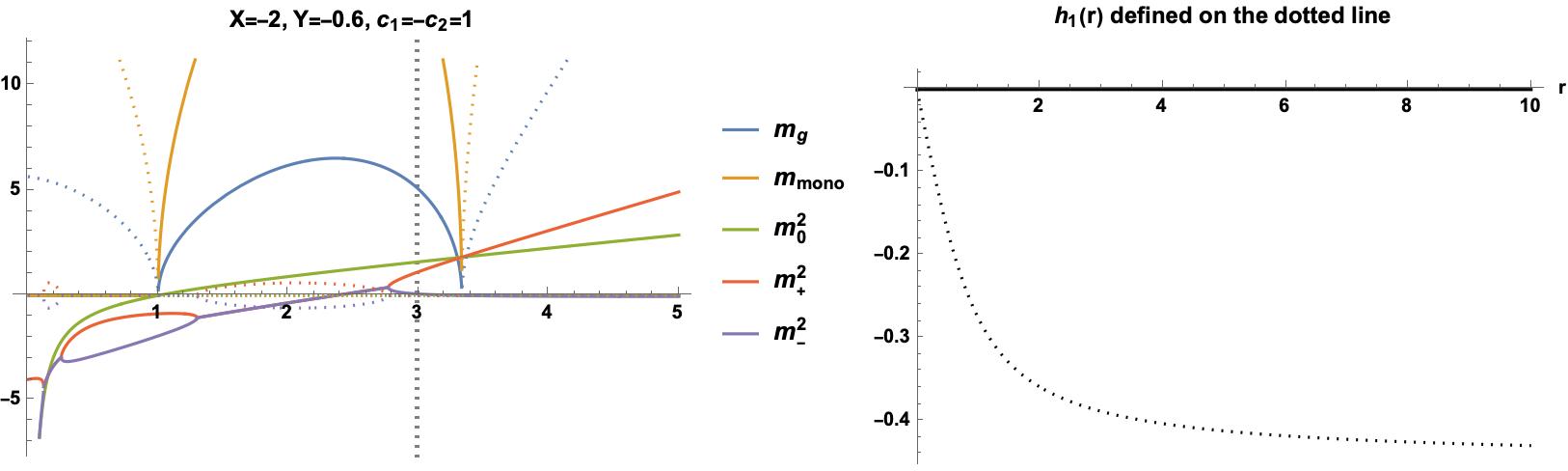}
                \caption{Both panels are plotted for $X=-2, Y=-0.6,c_1=-c_2=1,n=1, e=2$. The solid line represents the real part, and the dotted line represents the imaginary part of the masses.}
                \label{Figure: Ch3 Sec1 monopole vs gauge 6}
        \end{figure*}
        As one can see, intersection points of the physical regions of Higgs masses and monopole/gauge masses are non-trivial. In fact, they are bounded by two types of exceptional points. The first type is when two masses of Higgs particles coincide and form a complex conjugate pair. Such a point is known as an exceptional point where the mass matrix is non-diagonalisable, and the corresponding eigenvectors coincide. The second type is when the gauge and the monopole masses vanishes. Interestingly, this is where one of the Higgs masses also vanish. Since the mass matrix already has a zero eigenvalue, as the result of the spontaneous symmetry breaking, it seems the number of massless fields is increased. However, at this point, the mass matrix is also non-diagonalisable. Therefore one can not diagonalise the Hamiltonian to identify the field which corresponds to the extra massless fundamental field. Therefore this point is also an exceptional point. However, the eigenvalues do not become complex conjugate pairs beyond this point, and as one can see from figure \ref{Figure: Ch3 Sec1 monopole vs gauge vs Higgs} that one of the mass square $m_0^2$ become negative, and gauge and monopole masses become complex but with no conjugate pair. We dub such a point as zero exceptional point to distinguish from the standard exceptional point. 
    \subsection{Change in $\mathcal{CPT}$ symmetry and complex monopole solution}
\begin{figure*}[ht]
                \centering
                \includegraphics[scale=0.45]{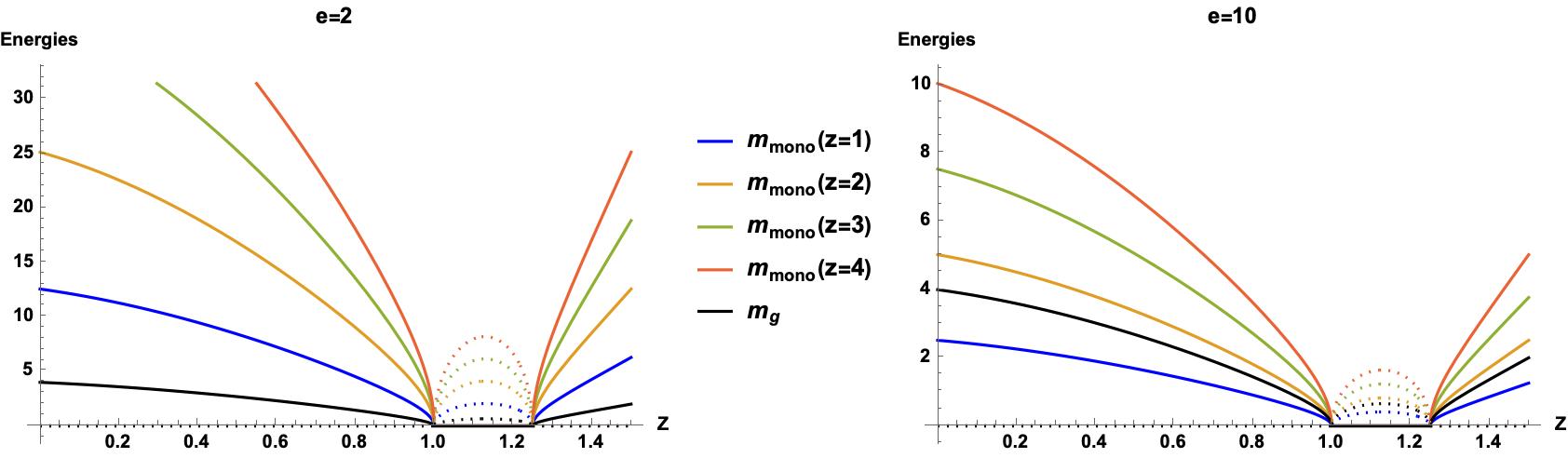}
                \caption{Monopole and gauge masses plotted for $X=1, Y=0.8, e=2, c_1=-c_2=1$. The solid line represents the real part, and the dotted line represents the imaginary part of the masses.}
                \label{Figure: Ch3 Sec1 monopole vs gauge 1}
            \end{figure*}

\begin{figure*}[ht]
                \centering
                \includegraphics[scale=0.5]{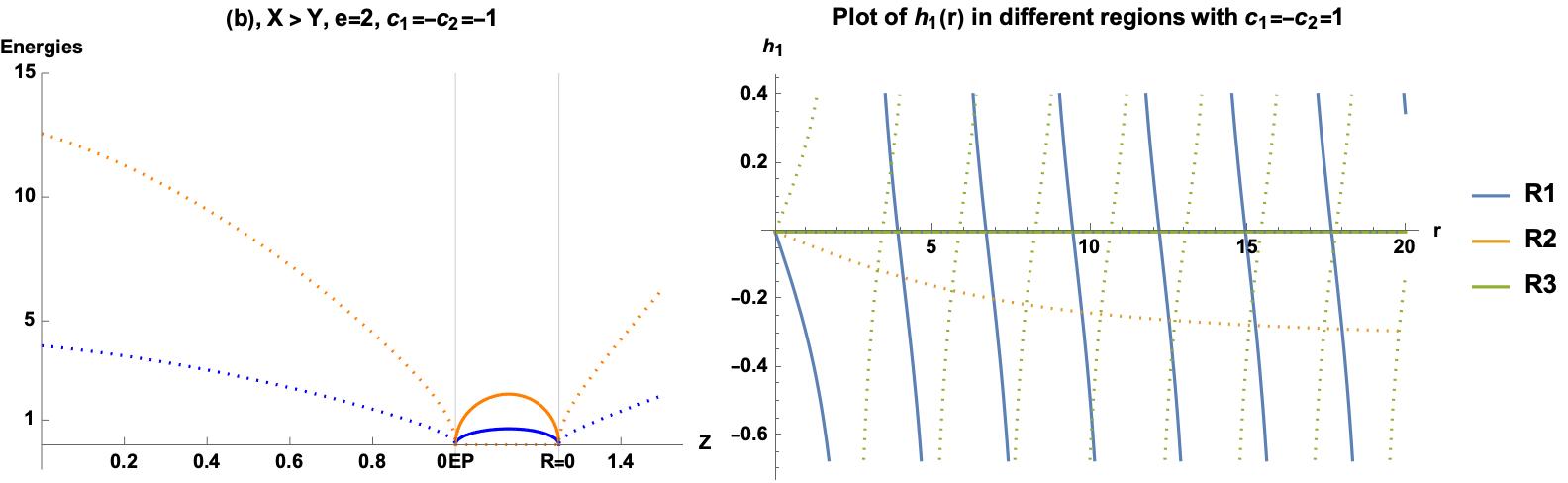}
                \caption{Both panels are plotted for $X=1, Y=1, n=1, e=2$. The solid line represents the real part, and the dotted line represents the imaginary part of the masses.}
                \label{Figure: Ch3 Sec1 monopole vs gauge 3}
        \end{figure*}

            We begin by introducing the useful quantities $m_1^2/g \equiv X, \mu^2/g\equiv Y ,\mu^2/m_2^2\equiv Z$. The gauge mass, monopole mass and monopole solutions can be rewritten in terms of these quantities 
            \begin{eqnarray}
                m_g = e R \sqrt{1-Z^2} ,~ m_\text{mono} = \frac{8|n|\pi R}{e}\sqrt{1-Z^2},\label{Equation: Ch3 Sec1 gauge and monopole masses}
            \end{eqnarray}
            \small
            \begin{eqnarray}
               \!\!\!\! h_1^\pm (r) \!\!\!\!&=& \!\!\!\!\pm \frac{\text{sign}(n)}{\sqrt{1-Z^2}}\big[R\sqrt{1-Z^2}\text{coth}(\hat{r})-\frac{1}{e r}\big],\label{Equation: Ch3 Sec1 profile function 1}\\
                \!\!\!\!h_2^\pm (r) \!\!\!\!&=& \!\!\!\!\mp \frac{\text{sign}(n) c_2 c_3 Z}{\sqrt{1-Z^2}}\left[R\sqrt{1-Z^2}\text{coth}(\hat{r})-\frac{1}{e r}\right],~~~~\label{Equation: Ch3 Sec1 profile function 2}
            \end{eqnarray}
            \normalsize
            where $R^2 = 4 (c_2 Z Y + c_1 X)$ and $\hat{r}=e R \sqrt{1-Z^2}r$.
            The monopole masses are plotted against the gauge mass for fixed parameters with $n\in\{1,2,3,4\}$ in figure \ref{Figure: Ch3 Sec1 monopole vs gauge 1} with weak and strong couplings $e=2, e=10$. Notice that the gauge mass is smaller than any of the monopole masses for weak coupling, but when $e$ is large enough, some of the monopole masses can become smaller than the gauge mass. This is clear by inspecting the monopole, and gauge mass in equation (\ref{Equation: Ch3 Sec1 gauge and monopole masses}) and two masses coincide when $e=\sqrt{8|n| \pi}$. Note that $n=0$ is not a monopole mass as it corresponds to the solution with zero winding number, which is topologically equivalent to the trivial solution. 
            \begin{figure*}[ht]
                \centering
                \includegraphics[scale=0.5]{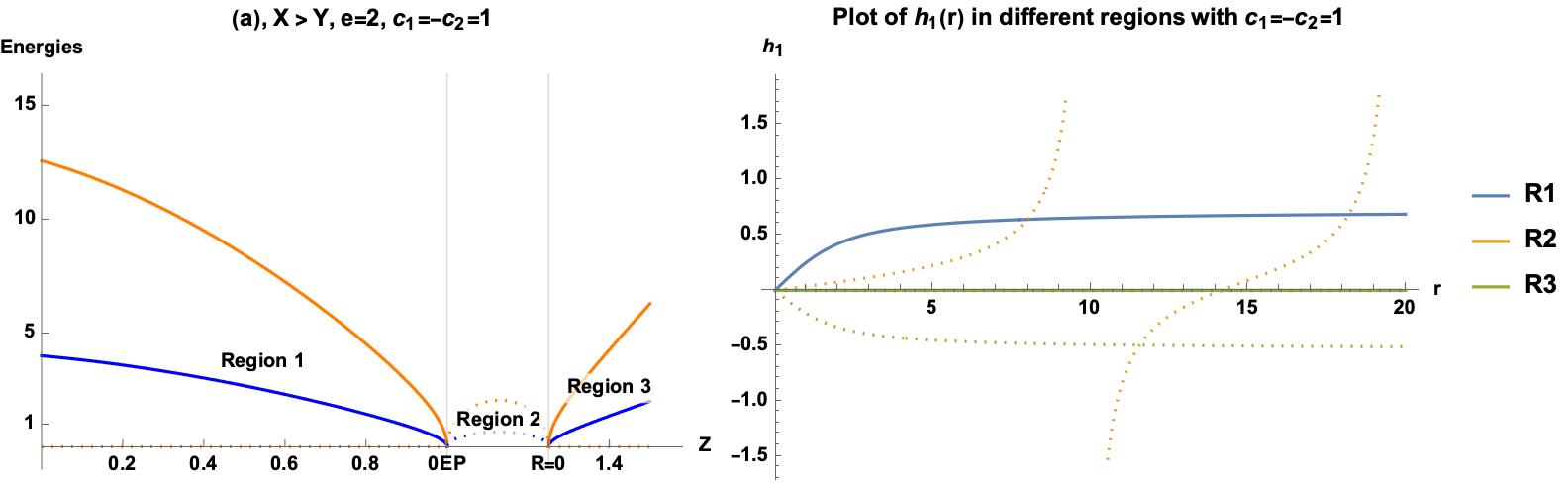}
                \caption{Both panels are plotted for $X=1, Y=0.8, n=1, e=2$. The solid line represents the real part, and the dotted line represents the imaginary part of the masses. Panel (a) shows the monopole and gauge masses against $Z\geq 0$, with vertical lines indicating the location of the boundaries of three regions. Panel (b) shows three profile function $h_1 (r)$ defined on each region indicated in panel (a).}
                \label{Figure: Ch3 Sec1 monopole vs gauge 2}
            \end{figure*} 
            From figure \ref{Figure: Ch3 Sec1 monopole vs gauge 1}, we also observe disconnected regions where both monopole and gauge masses become real to purely complex. A more detailed plot of this is shown in figure \ref{Figure: Ch3 Sec1 monopole vs gauge 2}. Region 2 is bounded by two points with lower bound $\mu^2 /m_2^2=1$ corresponding to the zero exceptional point where the vacuum manifolds stay finite (i.e. spontaneous symmetry breaking occur). However, the Higgs mechanism fails because the Hamiltonian is non-diagonalisable, as discussed in the previous section. The upper bounds correspond to the point where the vacuum manifold vanishes. Therefore, the spontaneous symmetry breaking does not occur, implying that the gauge fields do not acquire a mass through the Higgs mechanism, resulting in a massless gauge field. Most crucially, an interesting region (denoted by region 3 in figure \ref{Figure: Ch3 Sec1 monopole vs gauge 3}) reappears as one increases the value of $Z$. The profile function in region 3 is purely complex, which signals that this may lead to complex energies. However, as one can see from figure \ref{Figure: Ch3 Sec1 monopole vs gauge 2}, the energy is real. The reason for the real energy is that the conditions stated in the introduction hold. We will specify below the $\mathcal{CPT}$ symmetry responsible for the real value of the energy. Note that the profile function $h_2$ only differ from $h_1$ by some factor in front. Therefore we omitted it from the plot.

            Another physical region is when $c_1=-c_2 =-1$. The monopole and gauge masses for this case is plotted in figure \ref{Figure: Ch3 Sec1 monopole vs gauge 3}. We observe almost an identical plot from the figure \ref{Figure: Ch3 Sec1 monopole vs gauge 2} but with real and imaginary parts swapped. The profile functions also respect these changes as regions 1 and 3 no longer have a definite asymptotic value. The boundaries are unchanged, as one can see from the figure \ref{Figure: Ch3 Sec1 monopole vs gauge 3}. 
        
\begin{figure*}[ht]
                \centering
                \includegraphics[scale=0.5]{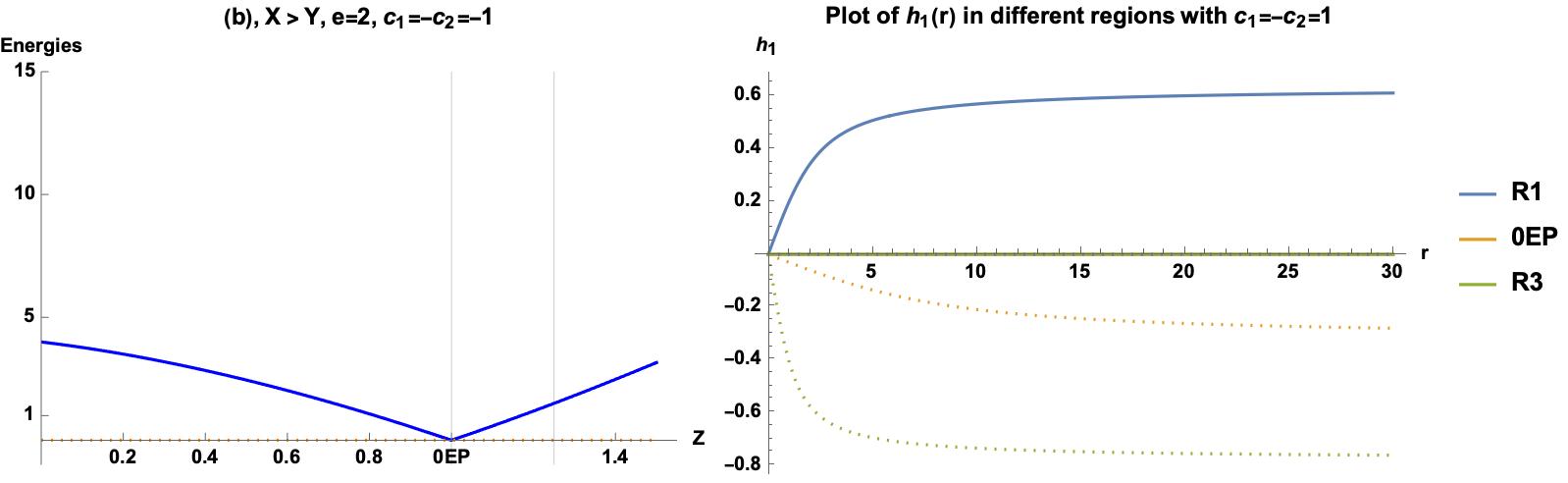}
                \caption{Both panels are plotted for $X=1, Y=1, n=1, e=2$. The solid line represents the real part, and the dotted line represents the imaginary part of the masses.}
                \label{Figure: Ch3 Sec1 monopole vs gauge 4}
        \end{figure*}

        Finally, there is an interesting parameter point $X=Y$ where region 2 vanishes (see figure \ref{Figure: Ch3 Sec1 monopole vs gauge 4}). The two boundaries $Z^2=1$ and $c_2 ZY +c_1 X =0 $ coincide when $X=Y$ and the zero exceptional point no longer exists because the spontaneous symmetry breaking does not occur in this case. 
        
        Next, let us explain the real value of the energies in different regions. First, to realise the conditions 1-3, stated in the introduction, we require the following transformations  
        \small
        \begin{equation*}
            \begin{array}{cc}
                h_2^\pm (r) \rightarrow  -h_2^\pm (r) ~,~~h_1^\pm (r) \rightarrow  h_1^\pm (r)   &  \text{in region 1} \\
                \text{No symmetry} &  \text{in region 2} \\
                h_2^\pm (r) \rightarrow  -\left(h_2^\pm (r)\right)^*~,~~h_1^\pm (r) \rightarrow  \left(h_1^\pm (r) \right)^*  &  \text{in region 3} 
            \end{array}.
        \end{equation*}
        \normalsize
        By using the explicit forms of the solutions (\ref{Equation: Ch3 Sec1 profile function 1}) and (\ref{Equation: Ch3 Sec1 profile function 2}). We can show that the above transformations satisfy condition 2 stated in the introduction, in regions 1
        \begin{eqnarray}
            h_2^\pm (r) \rightarrow  -h_2^\pm (r)=h_2^\mp (r) ~,~~h_1^\pm (r) \rightarrow  h_1^\pm (r) ,
        \end{eqnarray}
        and in region 3
        \begin{eqnarray}
            h_2^\pm (r) \rightarrow  -\left(h_2^\pm (r)\right)^*=h_2^\pm ~,~\\
            h_1^\pm (r) \rightarrow  \left(h_1^\pm (r) \right)^* =h_1^\mp . \nonumber
        \end{eqnarray}
        Notice that in regions 1 and 3, the $\mathcal{CPT}$ relates two distinct solutions in two different ways. For example, $h_2^\pm$ is mapped to $h_2^\mp$ in region 1, but it is mapped to itself in region 3.
        
        Finally, the condition 3 stated in the introduction is satisfied because the energy does not depend on the $\pm$ signs of the solutions. This explains the real energies of complex monopoles in region 3 and complex energy in region 2. Indeed, we observe the predicted behaviour in figure \ref{Figure: Ch3 Sec1 monopole vs gauge 2}. Region 2 is a hard barrier between two $\mathcal{CPT}$ symmetric regions where solutions are either real or purely imaginary. The same analysis can be carried out in the other physical region $c_1=-c_2 =-1$ where the symmetry is now 
        \begin{equation}
            \begin{array}{cc}
                \text{No symmetry}   &  \text{in region 1} \\
                \substack{h_2^\pm (r) \rightarrow  -\left(h_2^\pm (r)\right)^* \\ h_1^\pm (r) \rightarrow  \left(h_1^\pm (r) \right)^* }&  \text{in region 2} \\
                \text{No symmetry}   &  \text{in region 3} 
            \end{array} .
        \end{equation}
        We have observed that one can find a well-defined monopole solution in two disconnected regions. However, in the full theory where we include the Higgs particles, it is only one of the regions which are considered physical. This is because the Higgs mass $m_0^2$ is either positive or negative depending on which side of $Z^2=1$ it is defined. Because two disconnected regions are defined on either side of the zero exceptional point $Z^2=1$, the full physical region restricts one from moving region 1 to region 3 by changing $Z$. This is most clearly seen in figure \ref{Figure: Ch3 Sec1 monopole vs gauge vs Higgs} where the plot of $m_0^2$ (green line) becomes negative beyond the zero exceptional point. This may imply that the purely complex monopole solution we observed is not a possible solution of the theory. However, the purely complex solution can exist in the full physical region. An example of this is shown in the figure \ref{Figure: Ch3 Sec1 monopole vs gauge 6} where we observe that the profile function $h_1$ (therefore $h_2$) is purely complex, and the Higgs masses, gauge mass are all real and positive.
\section{Conclusions}
We have found the t'Hooft-Polyakov monopole solution (\ref{Equation: Ch3 Sec1 monopole solutions}) in the non-Hermitian theory by drawing an analogue from the standard procedure in the Hermitian theory. The monopole masses were plotted with the massive gauge and Higgs masses, where the physical region of the monopole masses coincided with that of the gauge mass. It was also observed that there are two distinct physical regions bounded by the zero exceptional point and the parameter limit where the vacuum manifold becomes trivial. The profile function (radial part of the monopole solution) is plotted in figures \ref{Figure: Ch3 Sec1 monopole vs gauge 2}, \ref{Figure: Ch3 Sec1 monopole vs gauge 3}, \ref{Figure: Ch3 Sec1 monopole vs gauge 4}, where it is real and purely complex in regions 1 and 3, respectively. Incidentally, the $\mathcal{CPT}$ symmetries of the solution are different in regions 1 and 3.

\section*{Acknowledgement}
TT is supported by EPSRC grant EP/W522351/1.

\bibliographystyle{phreport}
\bibliography{bib.bib}

\end{document}